\documentclass[10pt,a4paper,onecolumn]{article}
 \usepackage[utf8]{inputenc}
 \usepackage{amsmath}
 \usepackage{amsfonts}
 \usepackage{amssymb}
 \usepackage{graphicx}

\newcommand{\Ppump}{P_\mathrm{pump}}
\newcommand{\Pin}{P_\mathrm{in}}
\newcommand{\Pout}{P_\mathrm{out}}
\newcommand{\phiin}{\phi_\mathrm{in}}

\begin{document}

\title{Pulsed-pump phosphorus-doped fiber Raman amplifier around 1260 nm for applications in quantum non-linear optics}

\author{Eilon Poem$^*$, Artem Golenchenko, Omri Davidson,\\
Or Arenfrid, Ran Finkelstein, and Ofer Firstenberg}

\date{
    Physics of Complex Systems, Weizmann Institute of Science, Rehovot 7610001, Israel\\
    \bigskip
    \today
}

%\email{\authormark{*}eilon.poem@weizmann.ac.il} %% email address is required

%% To be edited by editor
% \dates{Compiled \today}

%\ociscodes{(140.3490) Lasers, distributed feedback; (060.2420) Fibers, polarization-maintaining;(060.3735) Fiber Bragg gratings.}

%% To be edited by editor
%\doi{}%\doi{\url{http://dx.doi.org/10.1364/XX.XX.XXXXXX}}

\maketitle

\begin{abstract}
We describe a fiber Raman amplifier for nanosecond and sub-nanosecond pulses centered around 1260 nm. The amplification takes place inside a 4.5-m-long polarization-maintaining phosphorus-doped fiber, pumped at 1080~nm by 3-ns-long pulses with a repetition rate of 200~kHz and up to 1.75~kW peak power. 
The input seed pulses are of sub-mW peak-power and minimal duration of 0.25~ns, carved off a continuous-wave laser with sub-MHz linewidth. We obtain linearly-polarized output pulses with peak-powers of up to 1.4~kW, corresponding to peak-power conversion efficiency of over 80\%. An ultrahigh small-signal-gain of 90~dB is achieved, and the signal-to-noise ratio 3~dB below the saturation power is above 20~dB. 
No significant temporal and spectral broadening is observed for output pulses up to 400~W peak power, and broadening at higher powers can be reduced by phase modulation of the seed pulse. Thus nearly-transform-limited pulses with peak power up to 1~kW are obtained. Finally, we demonstrate the generation of pulses with controllable frequency chirp, pulses with variable width, and double pulses. This amplifier is thus suitable for coherent control of narrow atomic resonances and especially for the fast and coherent excitation of rubidium atoms to Rydberg states. These abilities open the way towards several important applications in quantum non-linear optics.  
\end{abstract}

%\setboolean{displaycopyright}{false}

%\begin{document}

%\maketitle

\section{Introduction}
Coherent excitation of alkali atoms to Rydberg levels~\cite{Rydberg_Review_Saffman_2010} enables novel quantum-optics applications, such as the generation of non-classical light or the realization of deterministic photonic quantum gates~\cite{Ofer_JPB}. Initial studies of these applications have used ultra-cold atoms~\cite{Ofer13,Adams_PRL,Hofferberth_PRL}. However for practical and scalable implementations, it is preferable to operate with warm atoms around room temperature. For the typical wavelengths and volumes involved in Rydberg excitations, the atomic motion in warm  atomic vapors lead to decoherence of the excitation on a nanosecond time scale. Therefore, strong, sub-ns pulses are required for obtaining coherent excitations and for reaching the quantum nonlinear optics regime~\cite{Pfau_source}. The generation of such pulses requires dedicated light sources. Indeed, an ytterbium fiber amplifier system at 1010 nm, specifically designed for the excitation of rubidium atoms from the excited 6P manifold to Rydberg states was recently developed~\cite{Pfau_amp}. 

An alternative Rydberg excitation route in rubidium is a three-photon transition through the intermediate 5P and 5D levels. This route enables fast and efficient light storage for tens of nanoseconds on the 5D level~\cite{FLAME}. Additionally, as it involves three fields rather than two, it allows for partial mitigation of the Doppler dephasing~\cite{Adams_PRA,Ran_Power_Narrowing,Ran_Doppler_Compensation}. Finally, this route requires only infrared wavelengths, which are often preferable in terms of available laser power and compatibility with fiber-optics communication. 

Light between 1253 nm and 1270 nm is required for coupling the rubidium 5D level to Rydberg $n$P and $n$F levels (for principal quantum numbers $n>30$). Due to the relatively low transition dipole-moment to Rydberg levels, high peak-powers are required. For example, a full $\pi$-pulse transition from 5D$_{5/2}$ to 50P$_{3/2}$ using a Gaussian beam with a beam-waist radius of 30~$\mu$m and a Gaussian temporal pulse shape with a full width at half maximum (FWHM) of 0.25~ns requires a peak-power of about 400~W~\cite{ARC}. Longer pulses will expose the transition to Doppler dephasing (having a typical time scale of 1~ns), making it incoherent. Shorter pulses require even higher peak powers and may excite more than a single Rydberg level, as the typical separation between neighbouring high-$n$ levels is a few GHz~\cite{Rydberg_Review_Saffman_2010}. For these reasons, ps and fs pulses, which around this wavelength can be generated using an optical parametric oscillator~\cite{OPO} or a chromium:foresterite mode-locked laser~\cite{CrF}, are not suitable for this application. To the best of our knowledge, there are no existing laser systems that can generate strong enough, coherent, $\lesssim1$-ns-long pulses around 1260~nm. 

Here we present the development of a system designed for generating such pulses. We use an electro-optical modulator (EOM) to temporally carve the seed pulses out of a narrow-linewidth (sub-MHz) continuous-wave (cw) laser, and then amplify them using a custom, pulsed-pump, fiber Raman amplifier. 

Looking at common ways of light amplification, we see that while semiconductor laser-amplifier systems are commercially available for this wavelength range, they cannot go over 2~W or so~\cite{TA}. Doped-fiber amplifiers may have been able to go further, but unfortunately, there are no known doped gain fibers for this wavelength apart from bismuth-doped fibers~\cite{Bi_fibers}, which are not yet well understood and are not commercially available. In contrast, with a suitable pump laser, Raman amplification in commercially-available, undoped fibers can produce gain in this wavelength range, and indeed, such fiber Raman amplifiers are commercially available~\cite{Cont_RFA}. However, these amplifiers all use cw pumps, and therefore cannot go over a few tens of Watts, due mostly to overheating and Brillouin scattering. Additionally, as the nearest pump lasers currently available are based on ytterbium-doped fibers, with gain up to at most 1100 nm, multiple Stokes shifts of silica or germania glass are required to reach 1260 nm~\cite{Cont_RFA}. This may add unwanted noise and considerably complicate the system. 

Therefore, in order to reach peak powers of a few-hundred Watts with low noise in a compact design, we have developed a pulsed-pump Raman fiber amplifier based on a commercially-available phosphorus-doped fiber~\cite{Dianov2000_1,Dianov2000_2,Kim2000,Dianov2003,Xiong2003,Sim2004,Kurkov2007,Nagel2011,Kobtsev2015}. This fiber has an additional Raman gain band, due to P=O double bonds, at 40$\pm$0.3~THz (1330$\pm$10 cm$^{-1}$). This enables Raman amplification of 1260~nm signals using a commercially-available pump laser at 1080~nm, with a single Stokes shift. Using a pulsed pump both provides for high output peak powers with low average pump powers that are thermally manageable, and, with the pulses being being shorter than the acoustic phonon lifetime, ensures a high threshold for Brillouin scattering. 

Operating the system at 200~kHz, we successfully generate pulses of 0.25~ns FWHM with peak powers of up to 1.4~kW. Furthermore, 
up to output peak power of 400~W, we observe no significant temporal or spectral broadening, keeping the pulses transform-limited (TL). The signal-to-noise-ratio at this regime is higher than 20~dB. Above this peak power, spectral broadening becomes significant, due mostly to self-phase modulation~\cite{SPM1}. It can therefore be mitigated by pre-chirping the seed pulses, increasing the output power for which nearly-TL pulses could be achieved to 1~kW. Finally, we demonstrate the generation of amplified pulses with controllable duration up to 1~ns, pulses with controllable chirp rate, and double pulses of up to 1.6~ns temporal separation. 

\section{Setup}
\label{eesec:setup}
\subsection{Seed and amplifier}
 
\begin{figure}[htb]
\centering
\includegraphics[width=1\linewidth]{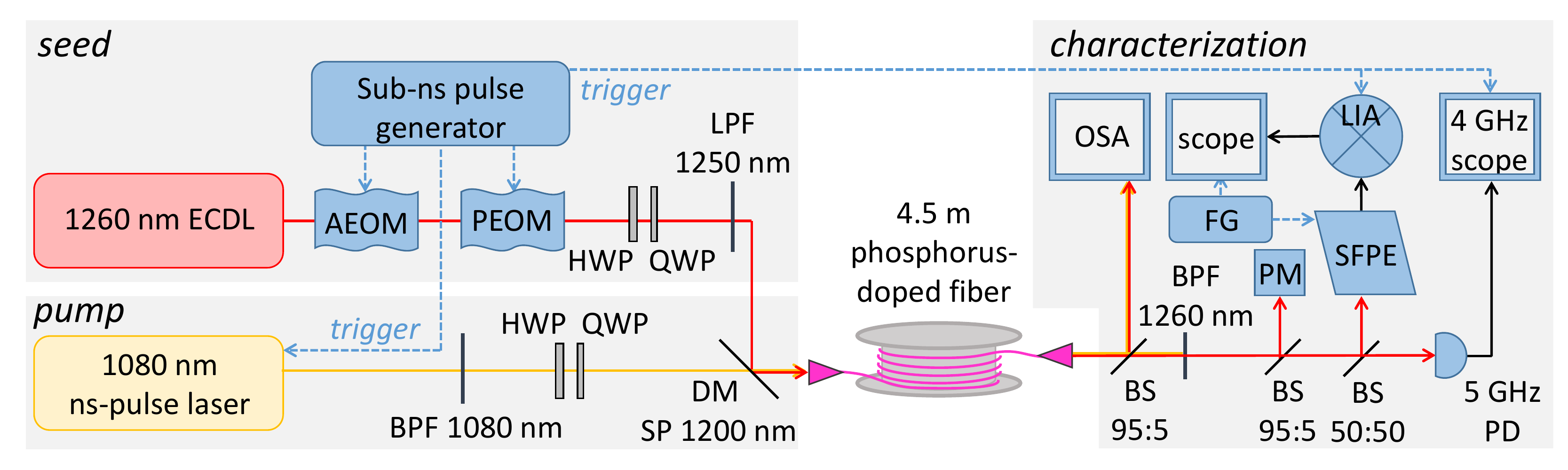}
\caption{\textbf{System.} Seed pulses are carved from an external-cavity diode laser (ECDL) by amplitude and phase electro-optical modulators (AEOM and PEOM) and enter the phosphorous-doped fiber together with the pump pulses. Pulse synchronization is ensured by an electronic pulse generator. The amplified pulses at the output of the phosphorus-doped fiber are spectrally and temporally characterized. HWP - half wave palte; QWP - quarter wave plate; DM - dichroic mirror; SP -short-pass; SPF - short-pass filter; LPF - long-pass filter; BPF - band-pass filter; BS - beam splitter; OSA - optical spectrum analyzer; PM - power meter; SFPE - scanning Fabri-Perot etalon; FG - function generator; LIA - lock-in amplifier; PD - photodiode.}
\label{fig:fig1}
\end{figure}

The system is schematically described in Fig.~\ref{fig:fig1}. A cw external-cavity diode laser (ECDL, Toptica DL-PRO with chip \#LD-1250-0100-AR-1) functions as a master oscillator. Custom seed pulses are generated by modulating this light using a 10~GHz amplitude EOM (AEOM, iXblu MX1300-LN-10) driven by a fast pulse generator (AT PG-1072). A phase EOM (PEOM, iXblu M1300-LN-10), driven by the same pulse generator, is used for controlling the temporal phase. 

A 1200 nm short-pass dichroic mirror (DM, Thorlabs DMSP1200T) is used for coupling the seed pulses into a 4.5-m-long polarization-maintaining phosphorus-doped fiber (PDF, FORC Photonics, PDF-5/125PM-P), where their amplification takes place. The maximal peak power of the seed pulses coupled into the PDF is 1.5~mW (80\% coupling efficiency), although eventually we use much weaker pulses. In order to allow for a high pumping power, the PDF is capped on both ends with $200$-$\mu$m-long core-less fiber pieces, thus increasing the damage threshold of the fiber facets. In addition, the input fiber coupler is temperature stabilized to 19$^\circ$C to allow for efficient heat extraction, and the space between the coupling lens and the fiber facet is purged by clean, dry nitrogen flow to prevent any humidity or dust accumulation on the fiber facet.

The pump light required for the amplification is at 1080 nm, which is one P=O Stokes shift to the blue of 1260 nm. This light is generated by an ytterbium-fiber-amplified pulsed diode laser (Cybel SpaceLight~1080) providing linearly polarized, 3-ns-long pulses with a spectral width of less than 0.1~nm centered around 1080~nm, at repetition rates between 200 kHz and 2 MHz. All results presented below are obtained for a repetition rate of 200 kHz. The wavelength can be varied between 1080 nm and 1084 nm by tuning the temperature of the diode laser. This tunability enabled Raman gain for signals between 1258 and 1267 nm (see below). The pump pulses are coupled to the PDF (70\% coupling efficiency) after passing through the DM, and co-propagate alongside the seed pulses. 
The pulse generator triggers the pump laser for synchronizing the pump and seed pulses. 

The polarizations of the seed and pump are aligned to the main polarization axis of the PDF by half- and quarter-wave plates. In order to protect the seed and pump lasers from back Raman scattering, a tilted 1300 nm long-pass filter (Thorlabs FELH1300) acting as a 1250 long-pass filter, and a band-pass filter (BPF, Semrock FF01-1055/70-25), are used on the seed and pump beams, respectively, before they combine at the DM.

At the fiber exit, a 3 nm band-pass filter around 1260 nm (Semrock FF01-1274/3-25) separates the amplified output pulses from residual pump light as well from light at wavelengths outside the 1260 nm region, which could be produced by Raman scattering of the silica glass (see below).

\subsection{Characterization setup}
In order to measure the broadband spectrum of the emitted light before the BPF, we use an optical spectrum analyzer (OSA, Ando AQ6317B). Then, after the BPF, we analyze the amplified pulses temporally and spectrally and use a power meter to measure their mean power. For the temporal analysis, we use a fast InGaAs photodiode (5 GHz, Thorlabs DET08CFC) and a fast scope (4 GHz, Keysight DSO9404A). For the spectral analysis, we use a scanning Fabri-P\'erot etalon (SFPE, Thorlabs SA210-8B) driven by a triangular wave from a function generator (FG, Keysight 33500B). The FWHM of the spectral response function of this SPFE is measured to be 90 MHz. The signal from the SFPE is locked-in to the repetition rate of the system by a fast lock-in amplifier equipped with an internal phase-lock-loop (LIA, Liquid Instruments Moku:Lab). The output of the LIA is displayed on a second scope (Keysight DSO-X 2004A), which is triggered by the FG. From the measured temporal and spectral intensity traces, the Gerchberg-Saxton (GS) algorithm~\cite{GerchSaxt} is used for retrieving the temporal phase $\phi(t)$ and extracting from it the instantaneous frequency $d\phi(t)/dt$.

\section{Results and discussion}
\subsection{Input pulses}
\begin{figure}[tb]
\centering
\includegraphics[width=1\linewidth]{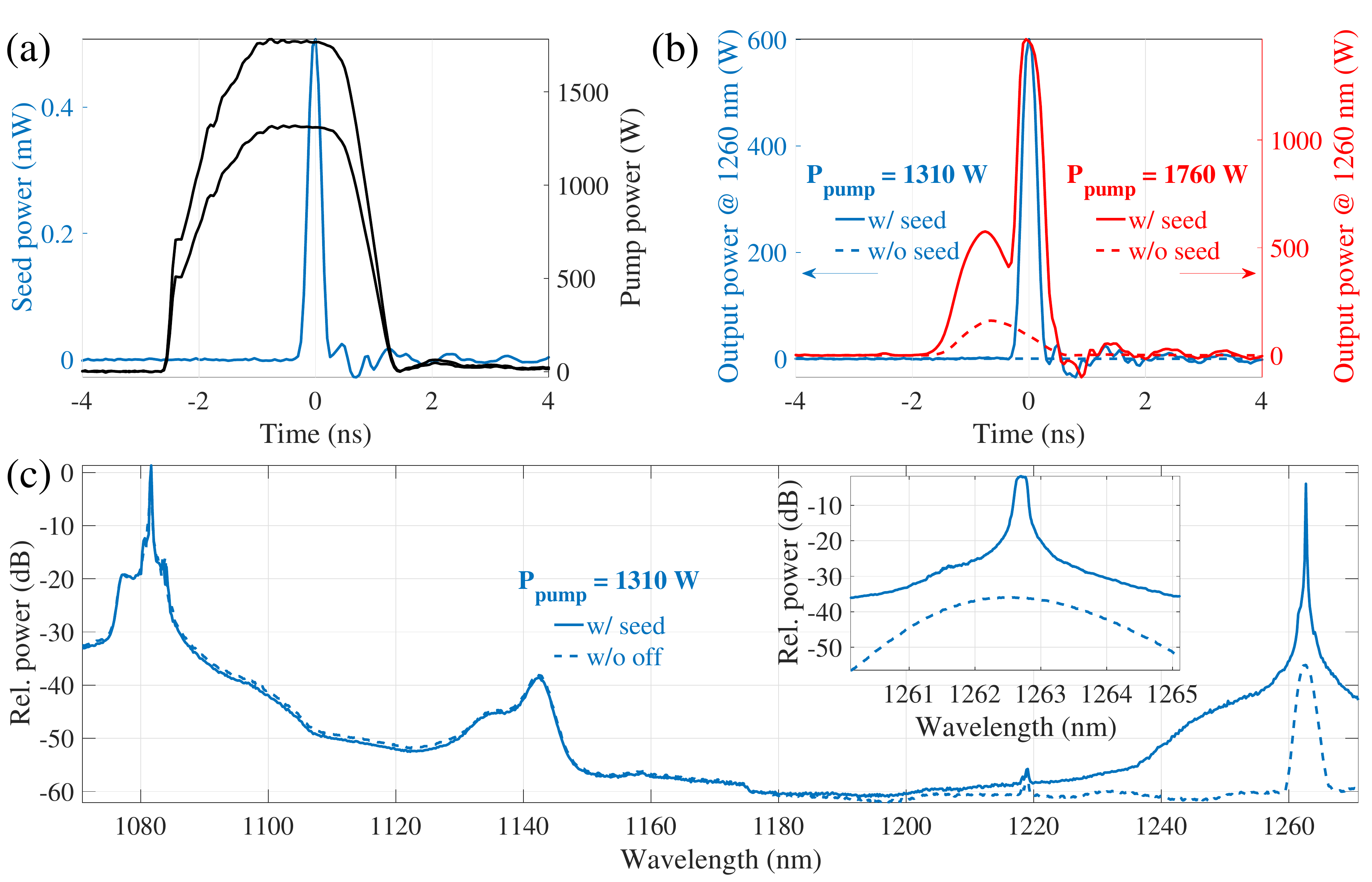}
\caption{\textbf{Input and output pulses.} (a) Time traces of a 250-ps-long seed pulse (blue, left axis) and pump pulses with intermediate ($\Ppump=1310~W$) and maximal ($\Ppump=1760~W$) peak powers (black, right axis). (b) Amplified 1260 nm pulses for the two pump pulses shown in (a) (solid lines) and the output noise in the absence of a seed (dashed lines). (c) Output spectra before the BPF with (solid) and without (dashed) the seed. Inset: Zoom in on the seed spectral region.}
\label{fig:fig2}
\end{figure}

A seed wavelength of 1262.6 nm (close to the 5D -- 43P resonance) is used for all results presented below, unless otherwise specified. A time-trace of the shortest seed pulse used in this study, with FWHM of 250 ns and no temporal phase modulation, is  shown in Fig.~\ref{fig:fig2}(a). 

The same figure also shows traces of pump pulses with two different peak powers, \mbox{$\Ppump=1310~W$} and \mbox{$\Ppump=1760~W$}. These are measured without the BPF, without any seed pulses, and in the polarization exhibiting minimal Raman scattering (45$^{\circ}$ off the fast axis of the PDF). The temporal shape does not considerably change with pump power.  

\subsection{Output pulses}
Figure~\ref{fig:fig2}(b) presents time traces of amplified seed pulses after the BPF, \emph{i.e.}, around 1260 nm, for the two pump powers shown in Fig.~\ref{fig:fig2}(a), as well as noise traces (no seed). While for high pump power, a significant background component emerges, it is negligible for the intermediate pump power. Here the seed peak power is fixed at a relatively high level of $\Pin=500~\mu$W; even at this input level, amplification of over 60 dB is measured.

In order to understand the source of the background emission, we measure the full spectra of at the output, before the BPF, using the OSA. Fig.~\ref{fig:fig2}(c) presents the spectra for the intermediate pump power, both with and without the seed. Background emission around 1260 nm appears due to self-stimulated first-order Raman scattering of P=O bonds. Some additional scattering at the first Stokes band of the silica glass (at around 1140 nm) appears as well, however it is well filtered out by the BPF.  

\subsection{Gain curves}
\begin{figure}[tb]
\centering
\includegraphics[width=1\linewidth]{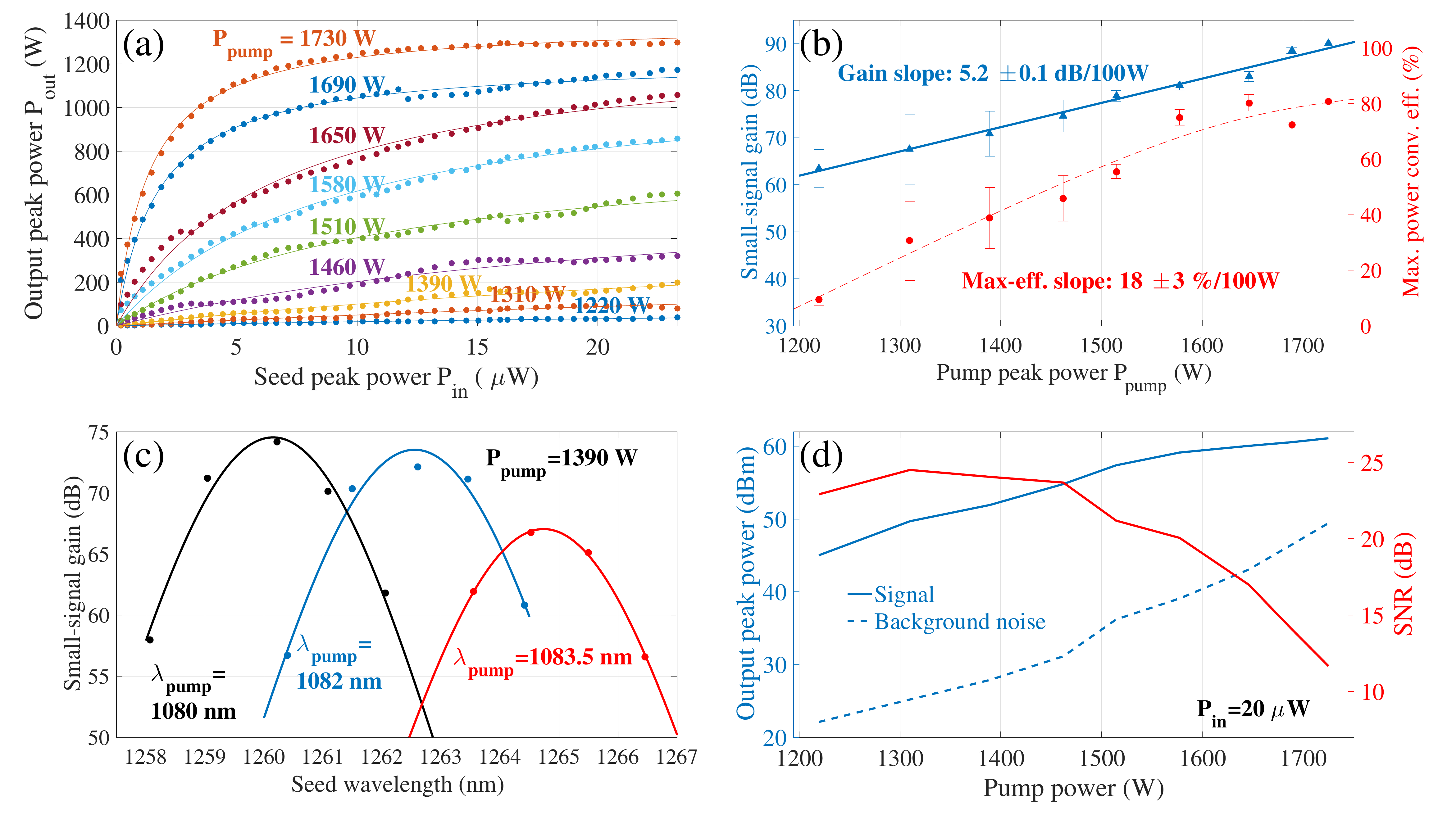}
\caption{\textbf{Amplifier characterization.} (a) Output peak power $\Pout$ versus seed peak power $\Pin$ for several pump peak powers $\Ppump$ (as labeled). The solid lines are fits to the function $\Pout=a \Pin/(\Pin+b)$. (b) The small signal gain and maximal power conversion efficiency, as extracted from the fits in (a), versus the $\Ppump$. The solid blue line is a linear fit, whereas the dashed red line is a guide to the eye. (c) Maximal small-signal gain at $\Ppump=1.39$~kW versus seed wavelength, for a few pump wavelengths (as indicated). The solid lines are Gaussian fits. (d) Left axis: output peak power (solid) and noise level (dashed) versus $\Ppump$. Right axis: Signal to noise ratio.}
\label{fig:fig3}
\end{figure}

Figure~\ref{fig:fig3}(a) presents the output peak power $\Pout$ versus the input peak power $\Pin$ for several pump peak powers $\Ppump$. A clear gain-saturation behaviour is observed. The solid lines are fits to a gain-saturation curve of the form $\Pout=a \Pin/(b+\Pin)$, where $a$ and $b$ are fit parameters. 

The small-signal gain in dB [10$\log_{10}(a/b)$] and the maximal peak-power conversion efficiency in percentage ($100a/P_{\mathrm{in}}$) are extracted from the fits and presented in Fig.~\ref{fig:fig3}(b). Clearly, both increase with the pump peak power, and the gain slope is 5.2~dB/100~W. At the highest $\Ppump$ (1.73 kW), the small signal gain is greater than 90~dB, and the peak output power is $\Pout=$1.4 kW, corresponding to 80\% peak-power conversion efficiency.

Additionally, we measured gain curves with $\Ppump=$1.39~kW for different seed and pump wavelengths. The extracted small-signal gain is presented in Fig.~\ref{fig:fig3}(c), showing that a gain above 50~dB is achievable in the spectral range of 1258-1267 nm. This range corresponds to transitions to all Rydberg levels laying between 37P and 57P~\cite{Rb_levels}.

Finally, figure \ref{fig:fig3}(d) presents the obtained $\Pout$ for $\Pin=20~\mu$W (solid blue line) and the background noise peak power (dashed black line), as well as the signal-to-noise ratio (SNR, red line). At this input power, SNR of over 20~dB is achieved for $\Ppump\le$1.5~kW, and even for the highest pump power used, the SNR is over 10~dB.

\subsection{Phase retrieval and control}
We have so far examined the pulse amplitudes and the background noise. As we are interested in transform-limited pulses, it is imperative to measure also their temporal phase and from it extract the instantaneous frequency. For this purpose, we simultaneously measure temporal traces and high-resolution spectral traces. We then run the GS phase-retrieval algorithm on the measured traces. 

\begin{figure}[tb]
\centering
\includegraphics[width=1\linewidth]{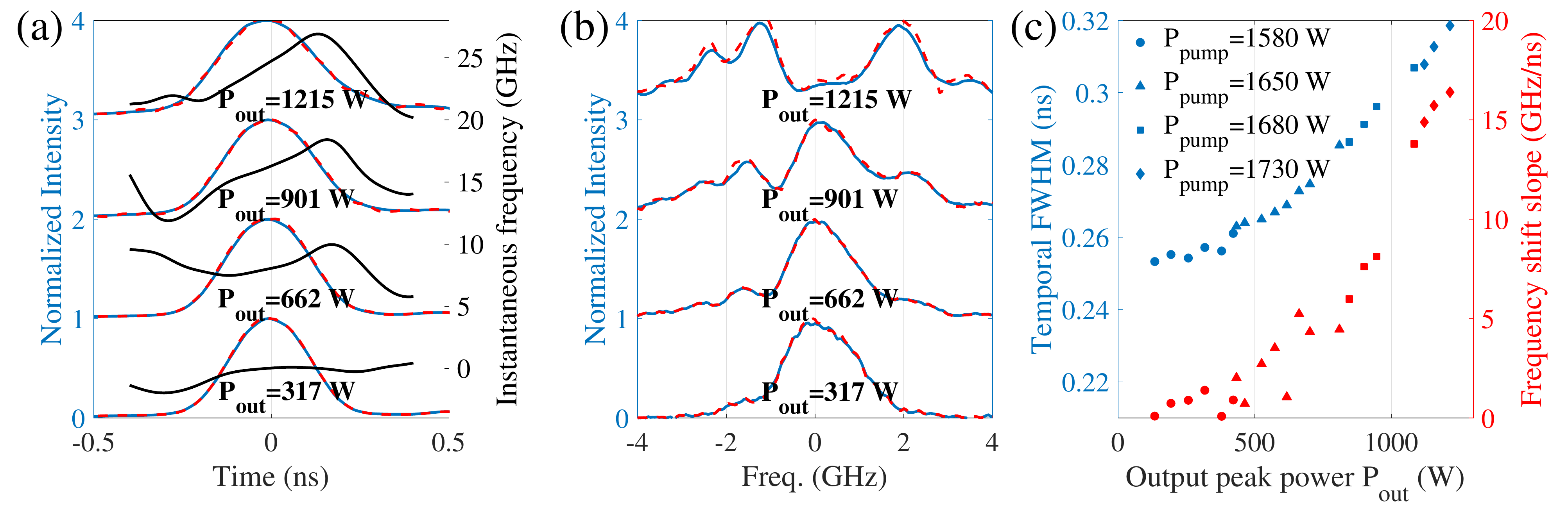}
\caption{\textbf{Phase retrieval.} (a) Left axis: measured (solid blue) and reconstructed (dashed red) normalized temporal traces of output pulses of various peak powers $\Pout$ (as labeled). Right axis: the corresponding extracted instantaneous frequency (the first derivative of the temporal phase). The traces are shifted vertically for clarity. (b) Measured (solid blue) and reconstructed (dashed red) normalized spectral traces of the same output pulses. The traces are shifted vertically for clarity. (c) Measured temporal FWHM (left axis) and chirp rate (the first derivative of the instantaneous frequency at the pulse center $t=0$) versus $\Pout$. The different symbols represent different pump peak powers (see legends).}
\label{fig:fig4}
\end{figure}

Figure~\ref{fig:fig4}(a) presents, for several $\Pout$, the measured temporal intensity traces (solid blue), the corresponding GS-reconstructed temporal traces (dashed red), and the corresponding extracted instantaneous frequency shifts (black). The latter are the first temporal derivative $d\phi(t)/dt$ of the phase $\phi(t)$ retreived by the GS reconstruction. Figure~\ref{fig:fig4}(b) presents the corresponding spectra from direct measurements and from the GS reconstruction. Using polynomial fits to $d\phi(t)/dt$, we extract the rate of frequency change (the chirp rate) at the pulse center for every pair of pump and seed peak powers. These chirp rates are presented versus the output peak power $\Pout$ in Fig.~\ref{fig:fig4}(c) (red). If two or more pairs of pump and seed powers produce the same output peak power, only the smallest extracted chirp rate is presented. The different symbols represent different pump powers. Also presented are the temporal FWHM of the output pulses (blue). Clearly, up to $\Pout=400$~W, neither a significant temporal broadening nor a significant frequency chirp are observed. However, a significant increase in both parameters is observed for $\Pout>400$~W, including a significant distortion of the pulse spectrum. 
These can be explained by the onset of self-phase modulation in the Raman gain fiber~\cite{SPM1}. %: the non-linear index of this fiber is XX, which implies that for fiber length of 4.5~m, significant phase will be imprinted above powers of about XXX W.

In order to compensate for the self-phase modulation and by that mitigate the temporal and spectral broadenings, we introduce an initial phase modulation to the seed pulses. To this end, we employ a PEOM between the seed and the amplifier. The temporal modulation is in the form of a pulse, identical in shape to the seed pulse. The peak magnitude $\phiin$ of this `phase pulse' is tuned within the range $0\le\phiin\le\pi$ until the measured spectral width at the output is minimized (the output phase is again retrieved using the GS algorithm). 
\begin{figure}[tb]
\centering
\includegraphics[width=1\linewidth]{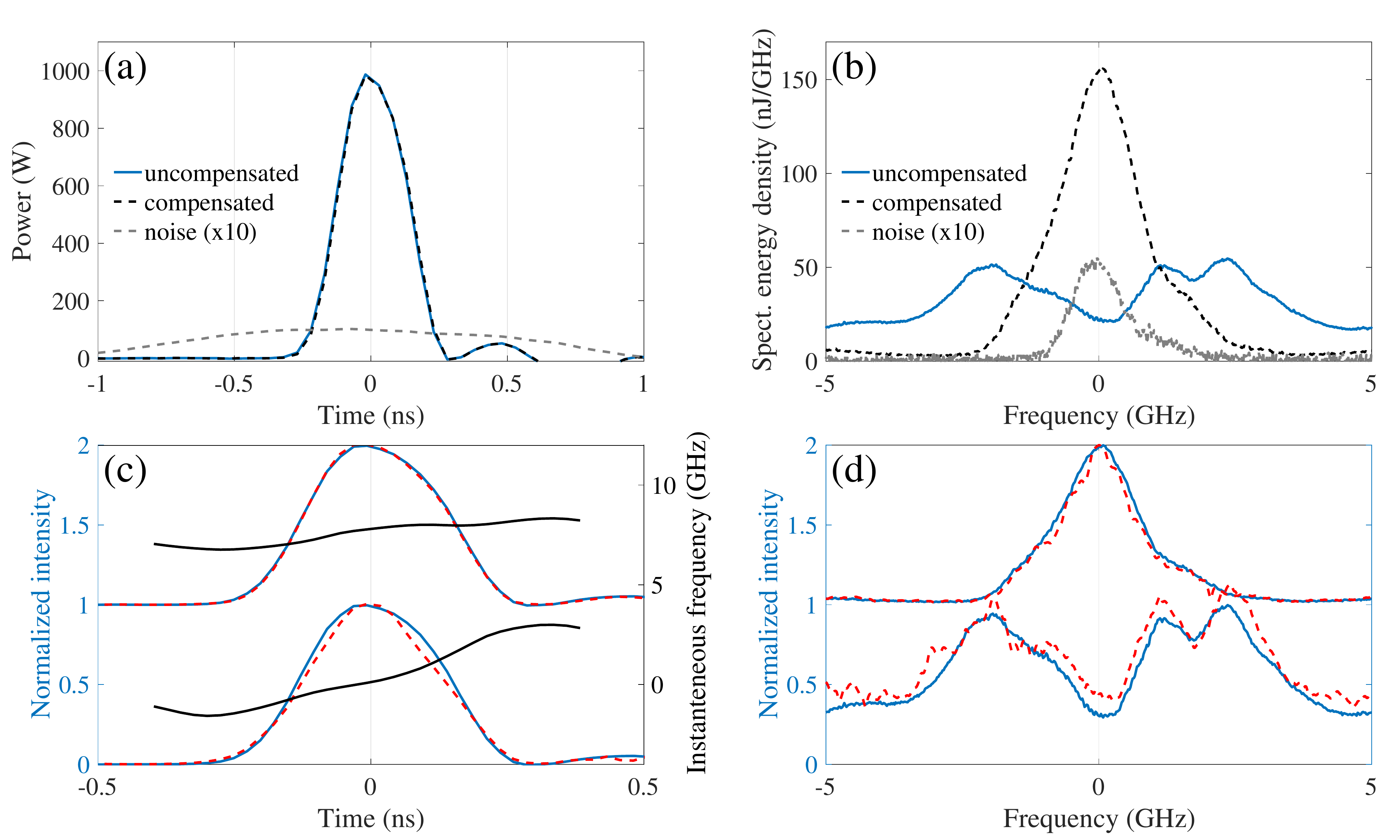}
\caption{\textbf{Phase compensation.} (a) Temporal and (b) spectral traces of output pulses without (dash black) and with (solid blue) phase compensation. The dashed gray line presents the background without seed. (c) Temporal and (d) spectral background-subtracted traces without compensation (bottom, solid blue) and with it (top, solid blue), together with their GS-reconstructed amplitudes (dashed red). The instantaneous frequency extracted from the GS reconstruction is presented by the solid black lines in (c) (right axis). The traces are shifted vertically for clarity. Here $\Ppump=1.6$~kW and $\Pin=100~\mu$W.}
\label{fig:fig5}
\end{figure}

Figures~\ref{fig:fig5}(a) and \ref{fig:fig5}(b) present the temporal and spectral traces of the output pulses both without and with phase compensation, as well as the noise background. While the temporal shape is nearly unchanged, the spectrum narrows down considerably. Figures~\ref{fig:fig5}(c) and \ref{fig:fig5}(d) present the GS reconstruction of the temporal and spectral traces without (bottom) and with (top) compensation. The black lines in Fig.~\ref{fig:fig5}(c) present the extracted instantenous frequency. The compensation clearly narrows down the spectrum and considerably reduces the frequency chirp during the pulse. 

\begin{figure}[tb]
\centering
\includegraphics[width=1\linewidth]{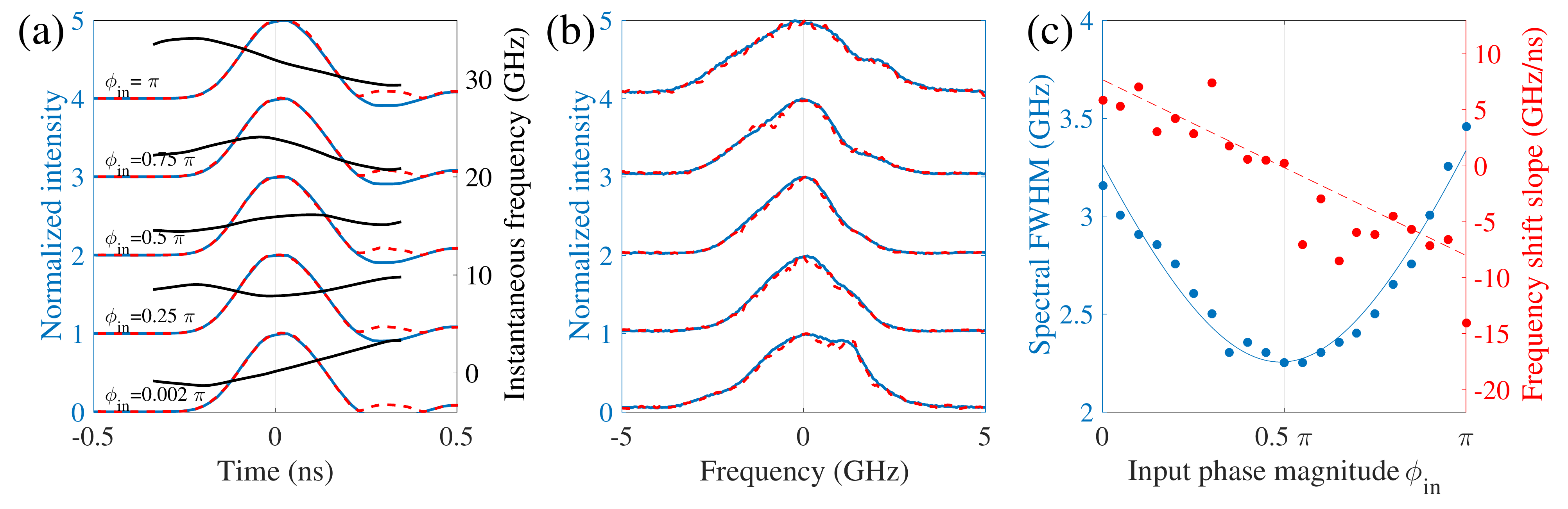}
\caption{\textbf{Phase control.} (a) Temporal and (b) spectral traces of output pulses for different magnitudes $\phiin$ of the input phase, as indicated. Measured data (solid blue lines) and the GS-reconstructed traces (dashed red lines) are presented, as well as the instantaneous frequency extracted by the GS reconstruction [black lines in (a)]. The traces are shifted vertically for clarity. (c) Spectral width (FWHM, blue circles) versus the input phase magnitude $\phiin$. The solid blue line is a fit to the expected hyperbolic dependence (see text). The chirp rate extracted from the GS reconstruction (red dots) agrees with the expected rate from the measured spectral widths (dashed line). Here $P_{\mathrm{pump}}=1.6$~kW, and $\Pin$ was tuned such that $\Pout=570$~W.}
\label{fig:fig6}
\end{figure}

Furthermore, the chirp rate could be controlled by tuning $\phiin$, such that both positively and negatively chirped pulses could be generated. This is shown for $\Pout=570$~W in Fig.~\ref{fig:fig6}. Figure~\ref{fig:fig6}(a) presents the measured and GS-reconstructed temporal traces, as well as the instantaneous frequency. Figure~\ref{fig:fig6}(b) presents the measured and GS-reconstructed spectral traces. The results are summarized in Fig.~\ref{fig:fig6}(c), where the spectral widths (left axis, blue) and the obtained chirp rates (right axis, red) are presented versus $\phiin$. The solid line is a fit to the expected hyperbolic dependence of the width $w$ on the phase magnitude $\phiin$, $w(\phi)=w_0\sqrt{1+(\phiin-\phi_{\mathrm{in,}0})^2/\eta^2}$, where $w_0$ is the minimal spectral FWHM, and $\phi_{\mathrm{in,}0}$ is the phase amplitude for which maximal compensation of the self-phase modulation occurs. We allow the temporal width of the phase pulse to deviate from that of the amplitude pulse by introducing the scale parameter $\eta$. Using this fit and the measured pulse duration (temporal FWHM) $\tau=0.245$~ns, the chirp rate $R$ can be expressed as $R(\phiin)=-2\ln{2}[(\phiin-\phi_{\mathrm{in,}0})/\eta]/(\pi \tau^2)$. This prediction is plotted by the dashed red line and agrees well with the GS-extracted chirp rates.

\subsection{Variable pulse shapes}
\begin{figure}[tb]
\centering
\includegraphics[width=1\linewidth]{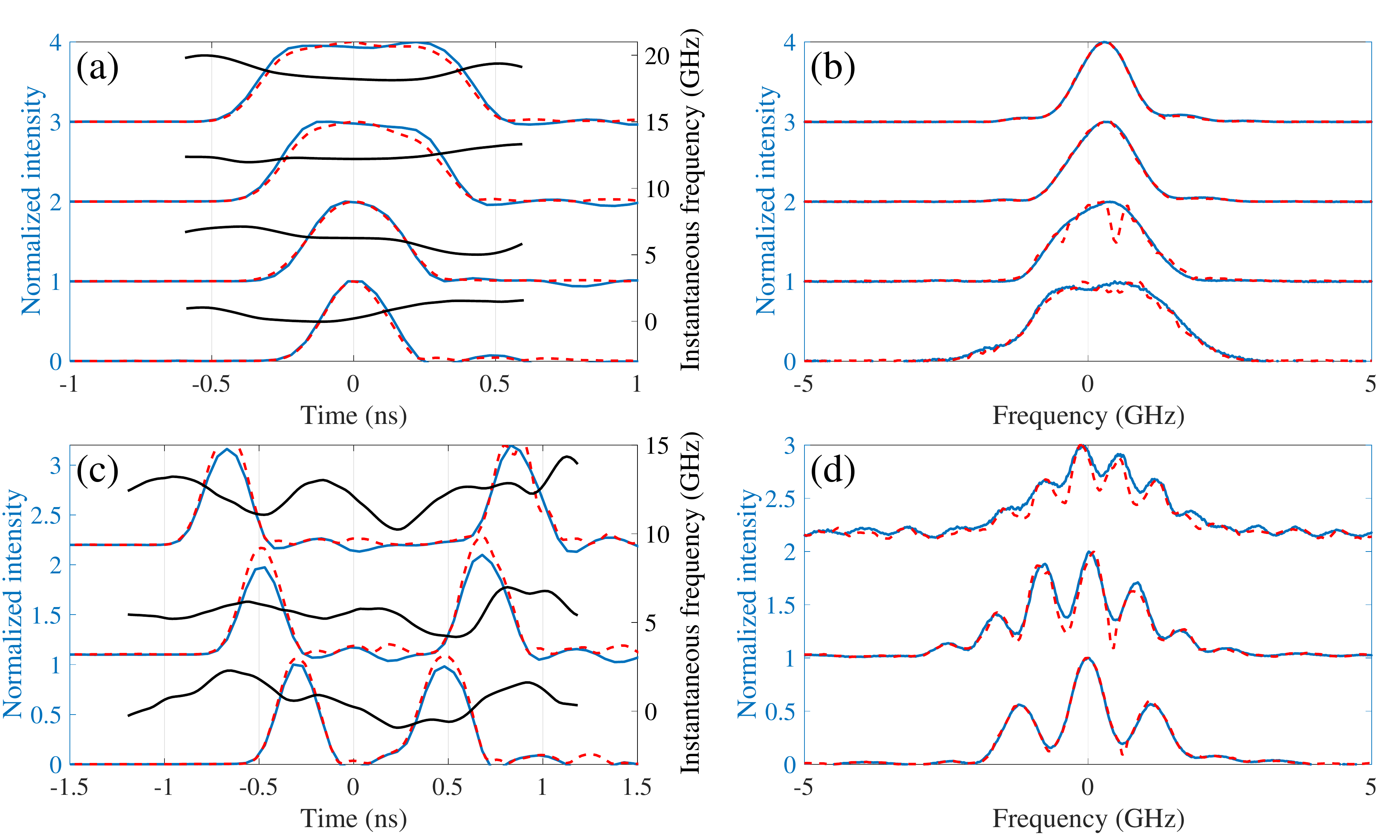}
\caption{\textbf{Variable pulse shapes}. Color coding is as in Figs.~\ref{fig:fig6}(a) and \ref{fig:fig6}(b).  (a) Temporal and (b) spectral traces of output pulses with variable duration. The output peak power is $\Pout=365$~W, and no phase compensation is employed. The traces are shifted vertically for clarity. (c) Temporal and (d) spectral traces of output double-pulses with variable temporal separations. The different line types represent the same parameters as in (a) and (b). Here $\Pout=500$~W, and partial phase compensation is employed (see text). The fringes in the spectra are an evidence of the coherence between the two pulses.}
\label{fig:fig7}
\end{figure}

In addition to single, 0.25-ns-long pulses, other pulse shapes can in principle be generated, for example by using an arbitrary waveform generator (AWG). Limited to the square pulse trains that our PG produces, we could only vary the seed pulse duration and the number of seed pulses within one pump pulse. Figure~\ref{fig:fig7}(a) presents output temporal traces of single pulses with various duration. We find that pulses with $\Pout=365$~W and temporal FWHM between 0.25~ns and 1~ns can be generated. Figure~\ref{fig:fig7}(b) presents the corresponding spectral traces. 
The GS-reconstructed traces and the extracted instantaneous frequency are presented as well. The GS reconstruction includes the 90-MHz spectral resolution of the SFPE. The nearly-constant instantaneous frequency traces show that the generated pulses are nearly transform limited. 

Figures~\ref{fig:fig7}(c) and \ref{fig:fig7}(d) present the corresponding temporal and spectral traces for trains of two 0.25-ns pulses with variable temporal separation, from 0.6 to 1.6~ns. As the temporal separation increases, the frequencies of the two amplified pulses drift apart (not shown). This results from the seed pulses approaching the outskirts of the pump pulse, where the pump intensity is not constant and varies in opposite directions for the two seed pulses [see Fig.~\ref{fig:fig2}(a)]. This, in turn, leads to a cross-phase modulation which causes opposite frequency shifts of the two amplified seed pulses\cite{XPM-shift,XPM-Raman}. 

To mitigate this, we apply a phase chirp to the seed pulses by introducing phase pulses that are slightly shifted from the amplitude pulses. While this enabled us to bring the frequencies of the two amplified pulses back to the same value, it also added some distortions to the pulses. This is evident in the GS-extracted instantaneous frequency [Fig.~\ref{fig:fig7}(c)]. 
Nevertheless, spectral fringes are still visible, even up to 1.6~ns separation, showing that the two pulses in the train are mutually coherent. The reduced fringe visibility can be explained by the limited resolution of the SFPE, as can be seen by the GS-reconstructed spectra, which take it into account.

\subsection{Possible improvements}
While the developed system achieves the goals of 1kW TL pulses with high SNR, it could be further improved on a few fronts.

First, the maximum power is currently not limited by the available pump power (which can go up to 4.2~kW in the PDF). The current limitation is the saturation of the first Stokes shift, as seen by the saturation of the conversion efficiency in Fig.~\ref{fig:fig3}(b), and the onset of the second Stokes shift. By using a shorter fiber, this point of saturation can be moved to higher pump powers, leading to higher output powers. Note that the noise level, the self-phase modulation, and the cross-phase modulation would stay roughly the same, as all scale as the product of the fiber length and the pump power~\cite{XPM-Raman}. 

Second, by employing stronger phase modulation with larger magnitudes $\phiin$, larger amplification-induced phases could be compensated for, further increasing the power of a TL pulse. In addition, using an AWG instead of the PG would enable the generation of arbitrary pulse shapes, as well as much better compensation of phase distortions.

Third, by using flatter pump pulses, cross-phase-modulation-induced frequency shifts could be completely avoided, and wider pulses or pulse trains could be more easily generated. Moreover, the wavelnegth range can be widen by using different diodes in the pump laser. This will make the amplifier compatible with transitions to Rydberg levels with principal quantum numbers $n<37$ and $n>57$.

Finally, the SNR could be further increased by using a narrower BPF at the output. BPFs with 50~GHz passband and
over 80\% transmission are commercially available for this wavelength.

\section{Conclusions}
We have built and tested a pulse-pumped Raman fiber amplifier for 1258-1267~nm. We observe the amplification of $250$~ps seed pulses with $\lesssim1$~mW peak power by up to 90~dB, to a maximal level of $\sim1.4$~kW. Up to 0.4~kW, the amplified pulses are nearly transform-limited. Above this power, significant self-phase modulation sets on. By pre-chirping the seed pulses, we can compensate for the self-phase modulation, thus increasing the maximal power of nearly transform-limited pulses to 1~kW. 

In addition, amplified pulses of controlled chirp rates and of various temporal widths, as well as pulse trains, can be generated. This opens the way towards coherent excitation of hot atoms to Rydberg states, with applications ranging from long and broadband light storage, through on-demand quantum light sources, to deterministic photonic quantum gates. 

Finally, other possible applications of the amplifier may include those based on the 1268.3~nm magnetic-dipole transition between triplet and singlet molecular oxygen. This transition is relevant for singlet oxygen level control~\cite{rev_sing_ox}, which is used, for example, in photodynamic therapy of cancer~\cite{ox_cancer}. A high spectral energy-density light source at this wavelength, as is within reach with the presented system, could prove beneficial for such applications.  

\section*{Funding Information}

Israel Science Foundation (ISF);
European Research Council (starting investigator grant QPHOTONICS
678674); the Pazy Foundation; the Minerva
Foundation with funding from the Federal German Ministry
for Education and Research; and the Laboratory in
Memory of Leon and Blacky Broder.

\section*{Acknowledgments}
The authors would like to thank Omer Ilday and Amiel Yishaaya for stimulating discussions, Yuval Rosenberg and Guy Hen for valuable technical assistance, and Jean-Marc Delavaux and Glen Williams of Cybel, LLC. for their help with the pump laser.\\

\flushleft{$^*$Corresponding author. For questions and comments please send an e-mail to \texttt{eilon.poem@weizmann.ac.il}}
%% Bibliography
%\bibliographystyle{amsplain}
%\bibliography{rfa}

\begin{thebibliography}{10}

\bibitem{Rydberg_Review_Saffman_2010}
M.~Saffman, T.~G. Walker, and K.~M\o{}lmer, {\em Quantum information with
  rydberg atoms}, Rev. Mod. Phys. {\bf 82} (2010), 2313--2363.

\bibitem{Ofer_JPB}
O~Firstenberg, C~S Adams, and S~Hofferberth, {\em Nonlinear quantum optics
  mediated by rydberg interactions}, Journal of Physics B: Atomic, Molecular
  and Optical Physics {\bf 49} (2016), 152003.

\bibitem{Ofer13}
Ofer Firstenberg, Thibault Peyronel, Q.-Y. Liang, Alexey~V. Gorshkov,
  Mikhail~D. Lukin, and Vladan Vuleti\'c, {\em Attractive photons in a quantum
  nonlinear medium}, Nature {\bf 502} (2013), 71.

\bibitem{Adams_PRL}
D.~Maxwell, D.~J. Szwer, D.~Paredes-Barato, H.~Busche, J.~D. Pritchard,
  A.~Gauguet, K.~J. Weatherill, M.~P.~A. Jones, and C.~S. Adams, {\em Storage
  and control of optical photons using rydberg polaritons}, Phys. Rev. Lett.
  {\bf 110} (2013), 103001.

\bibitem{Hofferberth_PRL}
H.~Gorniaczyk, C.~Tresp, J.~Schmidt, H.~Fedder, and S.~Hofferberth, {\em
  Single-photon transistor mediated by interstate rydberg interactions}, Phys.
  Rev. Lett. {\bf 113} (2014), 053601.

\bibitem{Pfau_source}
Fabian Ripka, Harald K{\"u}bler, Robert L{\"o}w, and Tilman Pfau, {\em A
  room-temperature single-photon source based on strongly interacting rydberg
  atoms}, Science {\bf 362} (2018), no.~6413, 446--449.

\bibitem{Pfau_amp}
Oliver de~Vries, Marco Pl\"{o}tner, Florian Christaller, Hao Zhang, Annika
  Belz, Benjamin Heinrich, Harald K\"{u}bler, Robert L\"{o}w, Tilman Pfau, Till
  Walbaum, Thomas Schreiber, and Andreas T\"{u}nnermann, {\em Highly customized
  1010~nm, ns-pulsed \uppercase{Y}b-doped fiber amplifier as a key tool for
  on-demand single-photon generation}, Opt. Express {\bf 28} (2020), no.~12,
  17362--17373.

\bibitem{FLAME}
Ran Finkelstein, Ohr Lahad, Ohad Michel, Eilon Poem, and Ofer Firstenberg, {\em
  Fast, noise-free memory for photon synchronization at room temperature},
  Science Adv. {\bf 4} (2018), eaap8598.

\bibitem{Adams_PRA}
N.~\ifmmode \check{S}\else \v{S}\fi{}ibali\ifmmode~\acute{c}\else \'{c}\fi{},
  J.~M. Kondo, C.~S. Adams, and K.~J. Weatherill, {\em Dressed-state
  electromagnetically induced transparency for light storage in uniform-phase
  spin waves}, Phys. Rev. A {\bf 94} (2016), 033840.

\bibitem{Ran_Power_Narrowing}
Ran Finkelstein, Ohr Lahad, Ohad Michel, Omri Davidson, Eilon Poem, and Ofer
  Firstenberg, {\em Power narrowing: counteracting doppler broadening in
  two-color transitions}, New J. Phys. {\bf 21} (2019), 103024.

\bibitem{Ran_Doppler_Compensation}
Ran Finkelstein, Ohr Lahad, Itsik Cohen, Omri Davidson, Shai Kiriati, Eilon
  Poem, and Ofer Firstenberg, {\em Optical protection of a collective state
  from inhomogeneous dephasing}, arXiv:2004.02295 (2020).

\bibitem{ARC}
N.~Šibalić, J.D. Pritchard, C.S. Adams, and K.J. Weatherill, {\em Arc: An
  open-source library for calculating properties of alkali rydberg atoms},
  Computer Physics Communications {\bf 220} (2017), 319 -- 331.

\bibitem{OPO}
L~Lefort, K~Puech, S.D Butterworth, G.W Ross, P.G.R Smith, D.C Hanna, and D.H
  Jundt, {\em Efficient, low-threshold synchronously-pumped parametric
  oscillation in periodically-poled lithium niobate over the 1.3 $\mu$m to 5.3
  $\mu$m range}, Optics Communications {\bf 152} (1998), no.~1, 55 -- 58.

\bibitem{CrF}
A.~A. Ivanov, A.~A. Voronin, A.~A. Lanin, D.~A. Sidorov-Biryukov, A.~B.
  Fedotov, and A.~M. Zheltikov, {\em Pulse-width-tunable 0.7~\uppercase{W}
  mode-locked \uppercase{C}r:forsterite laser}, Opt. Lett. {\bf 39} (2014),
  no.~2, 205--208.

\bibitem{TA}
D.~{Mehuys}, D.~F. {Welch}, and L.~{Goldberg}, {\em 2.0 \uppercase{W CW},
  diffraction-limited tapered amplifier with diode injection}, Electronics
  Letters {\bf 28} (1992), no.~21, 1944--1946.

\bibitem{Bi_fibers}
Young-Seok Seo, Yasushi Fujimoto, and Masahiro Nakatsuka, {\em Optical
  amplification in a bismuth-doped silica glass at 1300nm telecommunications
  window}, Optics Communications {\bf 266} (2006), no.~1, 169--171.

\bibitem{Cont_RFA}
S.~G. Grubb, T.~Erdogan, V.~Mizrahi, T.~Strasser, W.~Y. Cheung, W.~A. Reed,
  P.~J. Lemaire, A.~E. Miller, S.~G. Kosinski, G.~Nykolak, P.~C. Becker, and
  D.~W. Peckham, {\em 1.3 $\mu$m cascaded \uppercase{R}aman amplifier in
  germanosilicate fibers}, Optical Amplifiers and Their Applications, Optical
  Society of America, 1994, p.~PD3.

\bibitem{Dianov2000_1}
E.~M. Dianov, I.~A. Bufetov, M.~M. Bubnov, M.~V. Grekov, S.~A. Vasiliev, and
  O.~I. Medvedkov, {\em Three-cascaded 1407-nm \uppercase{R}aman laser based on
  phosphorus-doped silica fiber}, Opt. Lett. {\bf 25} (2000), no.~6, 402--404.

\bibitem{Dianov2000_2}
E.~M. Dianov and A.~M. Prokhorov, {\em Medium-power cw \uppercase{R}aman fiber
  lasers}, IEEE J. Sel. Top. Quantum Electron. {\bf 6} (2000), 1022--1028.

\bibitem{Kim2000}
N.~S. Kim, M.~Prabhu, C.~Li, J.~Song, and K.~Ueda, {\em 1239/1484 nm cascaded
  phosphosilicate \uppercase{R}aman fiber laser with cw output power of 1.36 w
  at 1484 nm pumped by cw yb-doped double-clad fiber laser at 1064 nm and
  spectral continuum generation}, Opt. Commun. {\bf 176} (2000), 219--222.

\bibitem{Dianov2003}
E.~M. Dianov, A.~S. Kurkov, O.~I. Medvedkov, V.~M. Paramonov, O.~N. Egorova,
  N.~Kurukitkoson, , and S.~K. Turitsyn, {\em \uppercase{R}aman fiber source
  for the 1.6--1.75 um spectral region}, Laser Phys. {\bf 13} (2003), 397--400.

\bibitem{Xiong2003}
Z.~Xiong, N.~Moore, Z.~G. Li, and G.~C. Lim, {\em 10~\uppercase{W}
  \uppercase{R}aman fiber lasers at 1248 nm using phosphosilicate fibers}, J.
  Lightwave Technol. {\bf 21} (2003), 2377--2381.

\bibitem{Sim2004}
S.~K. Sim, H.~C. Lim, L.~W. Lee, L.~C. Chia, R.~F. Wu, I.~Cristiani, M.~Rini,
  and V.~Degiorgio, {\em High power cascaded \uppercase{R}aman fibre laser
  using phosphosilicate fiber}, Electron. Lett. {\bf 40} (2004), 738--739.

\bibitem{Kurkov2007}
A.~S. Kurkov, V.~V. Dvoyrin, V.~M. Paramonov, O.~I. Medvedkov, and E.~M.
  Dianov, {\em All-fiber pulsed \uppercase{R}aman source based on
  \uppercase{Y}b:\uppercase{B}i fiber laser}, Laser Phys. Lett. {\bf 4} (2007),
  449--451.

\bibitem{Nagel2011}
J.~A. {Nagel}, V.~{Temyanko}, J.~{Dobler}, E.~M. {Dianov}, A.~S. {Biriukov},
  A.~A. {Sysoliatin}, R.~A. {Norwood}, and N.~{Peyghambarian}, {\em High-power
  narrow-linewidth continuous-wave \uppercase{R}aman amplifier at 1.27 $\mu$
  m}, IEEE Photonics Technology Letters {\bf 23} (2011), no.~9, 585--587.

\bibitem{Kobtsev2015}
Sergey Kobtsev, Sergey Kukarin, and Alexey Kokhanovskiy, {\em Synchronously
  pumped picosecond all-fibre \uppercase{R}aman laser based on phosphorus-doped
  silica fibre}, Opt. Express {\bf 23} (2015), no.~14, 18548--18553.

\bibitem{SPM1}
R.~C. Eckardt, C.~H. Lee, and J.~N. Bradford, {\em Effect of self-phase
  modulation on the evolution of picosecond pulses in a \uppercase{N}d:glass
  laser}, Opto-electronics {\bf 6} (1974), 67--85.

\bibitem{GerchSaxt}
R.~W. Gerchberg and W.~O. Saxton, {\em A practical algorithm for the
  determination of phase from image and diffraction plane pictures}, Optik {\bf
  35} (1972), 227.

\bibitem{Rb_levels}
J.~E. Sansonetti, {\em Wavelengths, transition probabilities, and energy levels
  for the spectra of rubidium (\uppercase{R}b \uppercase{I} through
  \uppercase{R}b \uppercase{XXXVII})}, J. Phys. Chem. Ref. Data {\bf 35}
  (2006), 301--421.

\bibitem{XPM-shift}
P.~L. Baldeck, R.~R. Alfano, and Govind~P. Agrawal, {\em Induced‐frequency
  shift of copropagating ultrafast optical pulses}, Applied Physics Letters
  {\bf 52} (1988), 1939--1941.

\bibitem{XPM-Raman}
A.~H\"{o}\"{o}k, D.~Anderson, and M.~Lisak, {\em Effects of cross-phase
  modulation and pump depletion on stokes pulse dynamics in stimulated raman
  scattering}, J. Opt. Soc. Am. B {\bf 6} (1989), 1851--1858.

\bibitem{rev_sing_ox}
Claude Schweitzer and Reinhard Schmidt, {\em Physical mechanisms of generation
  and deactivation of singlet oxygen}, Chem. Rev. {\bf 103} (2003), 1685--1757.

\bibitem{ox_cancer}
S.~G. Sokolovski, S.~A. Zolotovskaya, A.~Goltsov, C.~Pourreyron, A.~P. South,
  and E.~U. Rafailov, {\em Infrared laser pulse triggers increased singlet
  oxygen production in tumour cells}, Sci. Rep. {\bf 3} (2013), 3484.

\end{thebibliography}

\ifx\undefined\bysame
\newcommand{\bysame}{\leavevmode\hbox to3em{\hrulefill}\,}
\fi

\end{document}